\documentclass[letterpaper, 10 pt, conference]{ieeeconf}  

\IEEEoverridecommandlockouts                              

\usepackage{cite}
\usepackage{amsmath,amssymb,amsfonts}
\usepackage{algorithmic}
\usepackage{graphicx}
\usepackage{textcomp}
\usepackage{xcolor}

\usepackage{booktabs} 

\usepackage{multirow}
\usepackage{url}

\title{\LARGE \bf
Exploring Classical Piano Performance Generation with Expressive Music Variational AutoEncoder
}

\author{Jing Luo$^{1}$, Xinyu Yang$^{1}$, and Jie Wei$^{2}$
\thanks{$^{1}$Jing Luo and Xinyu Yang are with School of Computer Science and Technology, Xi'an Jiaotong University, Xi'an, 710049, China
        {\tt\small luojingl@stu.xjtu.edu.cn; yxyphd@mail.xjtu.edu.cn}}%
\thanks{$^{2}$Jie wei is with China Mobile Group Shaanxi Co., Ltd., Xi'an, China
        {\tt\small weijie@sn.chinamobile.com}}%
}

\begin{document}

\maketitle
\thispagestyle{empty}
\pagestyle{empty}

\begin{abstract}
The creativity of classical music arises not only from composers who craft the musical sheets but also from performers who interpret the static notations with expressive nuances. This paper addresses the challenge of generating classical piano performances from scratch, aiming to emulate the dual roles of composer and pianist in the creative process. We introduce the Expressive Compound Word (ECP) representation, which effectively captures both the metrical structure and expressive nuances of classical performances. Building on this, we propose the Expressive Music Variational AutoEncoder (XMVAE), a model featuring two branches: a Vector Quantized Variational AutoEncoder (VQ-VAE) branch that generates score-related content, representing the \emph{Composer}, and a vanilla VAE branch that produces expressive details, fulfilling the role of \emph{Pianist}. These branches are jointly trained with similar Seq2Seq architectures, leveraging a multiscale encoder to capture beat-level contextual information and an orthogonal Transformer decoder for efficient compound tokens decoding. Both objective and subjective evaluations demonstrate that XMVAE generates classical performances with superior musical quality compared to state-of-the-art models. Furthermore, pretraining the \emph{Composer} branch on extra musical score datasets contribute to a significant performance gain.
\end{abstract}

\section{Introduction}
In many musical genres, particularly Western classical traditions, performance is an indispensable element that transforms composers' abstract compositional ideas into engaging listening experiences through artistic decisions of performers \cite{cancino2018computational}. Recent advances in AI-Generated Content (AIGC) have sparked growing interest in automatically generating such expressive performances \cite{2020comprehensive}. This emerging field sits at the intersection of technological innovation and artistic creation, enabling novel applications ranging from virtual performers to enhanced creative tools for musicians.


Within piano performance generation research, two primary tasks have emerged: performance rendering \cite{jeong2019virtuosonet, zhang2024dexter} and performance generation from scratch \cite{oore2020time, muhamed2021symbolic}. The former refers to reproducing expressive parameters (e.g., tempo, timing, dynamics, articulation, and pedaling) for given musical scores, essentially simulating the role of pianists. The latter, more challenging task requires joint prediction of both score elements (e.g. notes, rhythm) and their expressive parameters, which constitutes the primary focus of this study. In this paper, we specifically address the two following critical challenges: music representation for expressive details of classical music and effective modeling frameworks.



Firstly, conventional approaches to classical piano music generation predominantly employ MIDI-like representations \cite{oore2020time}, based on the MIDI message protocol with musical token encoding via millisecond intervals. Although these representations faithfully represent most piano music, they omit metrical structures (e.g., beats) \cite{huang2020pop}, potentially leading to rhythmically inconsistent outputs. Subsequent studies have proposed alternative representations such as REMI \cite{huang2020pop}, Compound Word (CP) \cite{hsiao2021compound} and PerTok \cite{lenz2024pertok}. 
These methods are primarily adopted for music genres with stable tempos (e.g., popular music), but becomes insufficient for representing classical music performances where nuanced timing variations frequently occur even within individual beats.


Secondly, in physical music creation, there is a common division between the roles of composers and performers, particularly in classical music traditions. Composers define the abstract musical ideas in the score, specifying \emph{what} to play, while performers interpret the score, determining  \emph{how} to play it. Most prior research does not distinguish the roles between composer and performer \cite{huang2018music, muhamed2021symbolic, zhang2023disentangling}. In most cases, these methods generate the notes and their expressive parameters with same network architecture. These approaches place a heavy burden on the generative model, often resulting in suboptimal generation performance. 


To address the two challenges, we first introduce the Expressive Compound Word (ECP) representation, a novel method inspired by the Compound Word (CP) representation \cite{hsiao2021compound}. ECP faithfully encodes an expressive music into a sequence of compound tokens. At each time step, a single token comprises two distinct sets of sub-tokens: four score-related sub-tokens that define the notes and metrical structure, and four performance-related sub-tokens that specify detailed expressive parameters of each note. 

Then, we propose the Expressive Music Variational AutoEncoder (XMVAE), a novel Seq2Seq based framework designed to generate classical piano performances from scratch. 
XMVAE employs a two-branch Transformer-based architecture that metaphorically divides the task between a \emph{Composer} and a \emph{Pianist}. The \emph{Composer} branch, a VQ-VAE, operates at the note level to generate the detailed scores and outline the metrical structure. In contrast, the \emph{Pianist} branch, a vanilla VAE, functions at a higher, whole-song level. It interprets the generated score by learning a global representation of performance style.
These two branches are trained jointly, ensuring seamless integration between composition and the generation of expressive parameters. Specifically, we incorporate a multiscale encoder in the encoder blocks of both branches, enabling additional context learning at the beat level. Furthermore, we design orthogonal Transformer decoders for both branches to efficiently decode compound tokens across both the temporal axis and the sub-token axis.

We conducted both objective and subjective evaluations on classical piano performance datasets to validate the effectiveness of our proposed methods. Both evaluations consistently demonstrate that our approach outperforms the existing benchmark models across most metrics. Furthermore, through ablation studies, we observed that pretraining the \emph{Composer} branch of XMVAE on an extra musical score dataset significantly enhances overall performance.


\section{Related Work}
\label{2_related_work}
Prior studies on piano performance generation from scratch typically rely on symbolic format like MIDI rather than audio, as MIDI offers an editable and detailed set of performance parameters that can produce highly realistic musical results. 

The music representations in the most early studies follow the MIDI standard \cite{oore2020time, huang2018music, muhamed2021symbolic}, where temporal tokens are quantized into millisecond bins, leading to the exclusion of metrical structure. To address this, later studies proposed alternative music representations, such as REMI \cite{huang2020pop} and CP \cite{hsiao2021compound}. REMI and CP integrate explicit metrical grids to solve metrical structural modeling and enhance expressive dimensions through performance tokens including Velocity and Tempo. 
However, the Tempo tokens in both REMI and CP operate at the beat level rather than the note level, leading to their limited applications to capture the nuanced expressiveness of classical music, where timing often vary at the note level. 

The recently proposed PerTok \cite{lenz2024pertok} addresses performance nuances by introducing microshift tokens, which encode MIDI tick offsets from quantized note positions. However, as a REMI-based representation, PerTok still produces longer token sequences compared to CP-like formats. More critically, its design is focused on fixed-tempo modern genres \cite{lenz2024pertok}, which limits its scalability to classical music that often includes frequent tempo fluctuations and is performed in a rubato style. As a result, the quantized note positions in PerTok often misalign with note onsets defined in classical scores, risking potential distortion of the metrical structure.

Performance RNN \cite{oore2020time} and Music Transformer \cite{huang2018music} are two of the most influential early studies on performance generation from scratch. Both studies leveraged vanilla sequential models (RNN or Transformer) as backbone architecture. Transformer-GAN \cite{muhamed2021symbolic} further enhanced the quality of generated samples by incorporating Generative Adversarial Networks (GANs). PerfVAE \cite{zhang2023disentangling} explored the disentanglement of musical content and performance style in classical piano performances using a Variational Autoencoder (VAE) model. These studies don't design the specific modules to decode the score tokens and performance tokens separately, leading to suboptimal generated outputs. Recent study like Cadenza \cite{lenz2024pertok} introduced a two-stage framework for generating expressive music variations for popular music, which consists of Composer model creating music variations and Performer model infilling expressive details. However, the two models of Cadenza are trained independently and inferred sequentially, which can introduce potential sampling errors.

\section{Methods}
\label{3_methods}
\subsection{Expressive Compound Word Representation}
\label{3_ECP}
\begin{figure*}[!t]
    \begin{minipage}[b]{0.36\textwidth}
        \centering
        \includegraphics[width=1.1\textwidth]{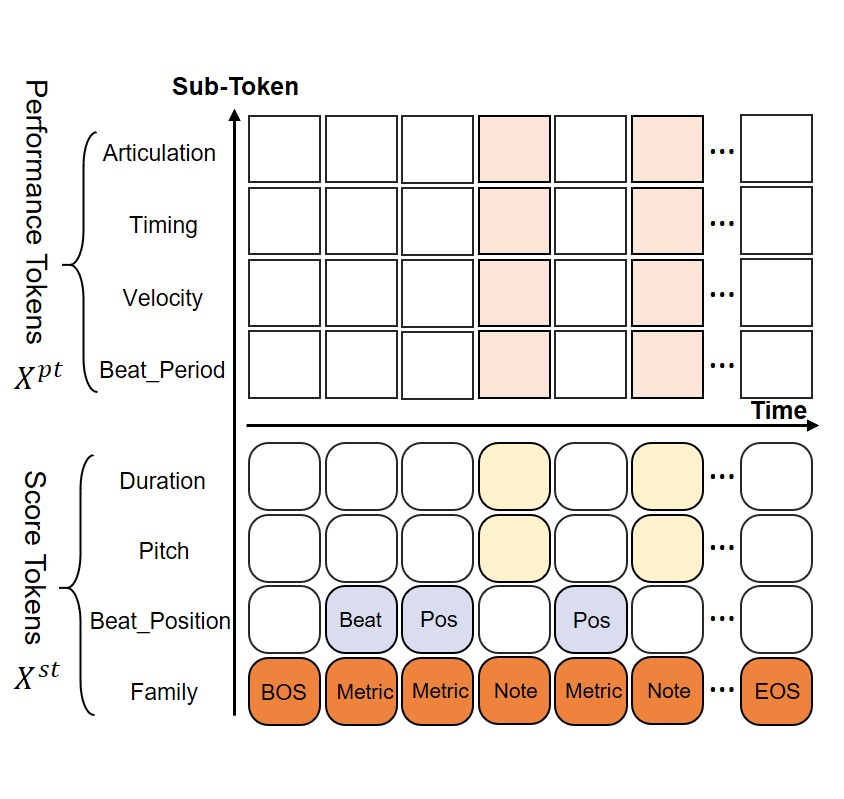}
        \caption{Expressive Compound Word Representation. We pad the missing token types with “ignore” tokens at each time step, denoted as squares with empty fillings.}
        \label{ecp_representation}
    \end{minipage} \quad \quad
    \begin{minipage}[b]{0.63\textwidth}
        \centering
        \includegraphics[width=0.9\textwidth]{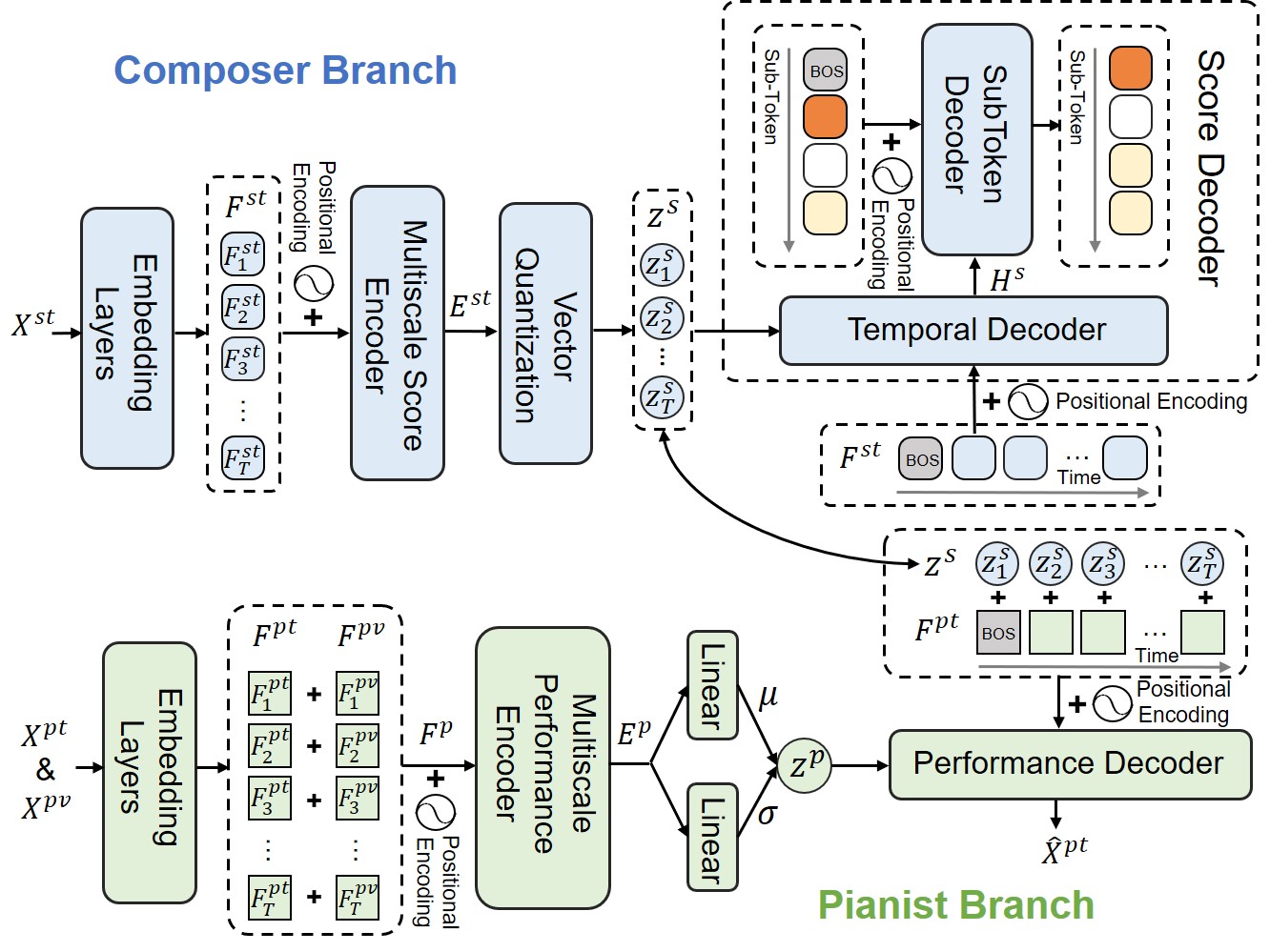}
        \caption{The overall architecture of XMVAE consists of two branches: the upper \emph{Composer} and the lower \emph{Pianist}. For clarity, only the detailed architecture of the Score Decoder in the \emph{Composer} branch is shown, as the Performance Decoder in the \emph{Pianist} branch shares the same structure.}
        \label{xmvae_overall}
    \end{minipage}
\end{figure*}

To accurately capture both the metrical structure and nuanced expressive details of classical music, we introduce the Expressive Compound Word (ECP) representation. This tokenizer uses beat as the basic metrical boundary and represents classical performance as a sequence of compound tokens, following the token organization of the original CP \cite{hsiao2021compound}. The compound token of ECP at each time step consists of four score sub-tokens and four performance sub-tokens as shown in Fig.~\ref{ecp_representation}.

\textbf{Score Tokens $X^{st}$:} Four tokens represent score features, covering the note information and metrical structure: \textbf{Family} specifies the type of compound token, which includes four categories: [Note], [Metric], [BOS], and [EOS], consistent with the original CP. \textbf{Beat\_Position} marks the beginning of a new beat or position within a beat. The temporal resolution of a beat is set to 24 in this paper. \textbf{Pitch} corresponds to the MIDI pitch number of piano, ranging from 21 to 108. \textbf{Duration} represents the abstract duration of a note in the score, encoded using 51 tokens with varying levels of quantization for both short and long durations following \cite{luo2024bandcontrolnet}.

\textbf{Performance Tokens $X^{pt}$:} Four tokens are quantized from the following four continuous real-valued expressive parameters $X^{pv}$ \cite{partitura_mec, zhang2024dexter}, which specify the detailed expressive details of every note appearing in the score: \textbf{Beat\_Period} denotes the ratio of the inter-onset-interval (IOI) between two consecutive notes in the performance and score. \textbf{Velocity} means the MIDI velocity number, ranging from 0 to 127. \textbf{Timing} represents the difference between the onset time of a note played in performance and that in the score. \textbf{Articulation} is the ratio of the actually played duration to the duration in the score.

These continuous expressive parameters are quantized into discrete values for event-based tokenization. Specifically, \textbf{Beat\_Period}, \textbf{Timing}, and \textbf{Articulation} are quantized into 161, 41, and 81 tokens, respectively, with logarithmic distribution. Meanwhile, \textbf{Velocity} is quantized into 32 evenly distributed tokens. Performance tokens only will available when \textbf{Family} token is [Note].

Overall, we represent a piano music as two groups token sequences: $X^{st}=\{\{x^{st}_{i,j}\}^4_{j=1}\}^{T}_{i=1} \in \mathbb{R}^{T \times 4}$ and $X^{pt}=\{\{x^{pt}_{i,j}\}^{4}_{j=1}\}^{T}_{i=1} \in \mathbb{R}^{T \times 4}$, where $T$ is the sequence length, $x^{st}_{i,j}$ and $x^{pt}_{i,j}$ are the $j^{th}$ sub-token of score and performance respectively at $i^{th}$ time step. Note that the discrete performance tokens $X^{pt}$ are quantized from their corresponding real-valued expressive parameters $X^{pv} \in \mathbb{R}^{T \times 4}$. 

\subsection{Expressive Music Variational AutoEncoder Model}
\label{XXXb}
We present our Expressive Music Variational AutoEncoder (XMVAE) as illustrated in Fig~\ref{xmvae_overall}, which consists of a VQ-VAE \cite{van2017neural} based \emph{Composer} branch for generating notes and outlining metrical structure, and a vanilla VAE \cite{kingma2013auto} based \emph{Pianist} branch for generating expressive parameters given the encoded VQ codes from \emph{Composer} branch as the score features. Both two branches leverage crafted multiscale encoder and orthogonal Transformer decoder. 


\subsubsection{Composer Branch}
Given the score token sequences $X^{st}=\{\{x^{st}_{i,j}\}^{4}_{j=1}\}^{T}_{i=1}$, we first compute the inputting embeddings $F^{st}=\{F^{st}_i\}^T_{i=1} \in \mathbb{R}^{T\times d}$ by summing the multiple sub-token embeddings at each time step, as follows:
\begin{equation}
    \label{eq_embedding}
    \begin{aligned}
        & F^{st}_{i} = \sum^4_{j=1} F^{st}_{i,j}, \\
        & F^{st}_{i,j} = \mathbf{TE}^{st}_{j}(x^{st}_{i,j}), \\
    \end{aligned}
\end{equation}
where $\mathbf{TE}$ represents the embedding layer for each sub-token, and $d$ denotes the embedding dimension. 

The embeddings $F^{st}$ are then passed directly to the Multiscale Score Encoder ($\mathbf{Enc^s}$), which integrates a standard self-attention module \cite{vaswani2017attention} and a beat-level self-attention module. The beat-level self-attention module restricts the attention range of each time step to its corresponding beat, enabling effective learning of contextual features at the beat level. After processing through the multiscale encoder, we obtain the output $E^{st}=\{E^{st}_i\}^T_{i=1}$ as follow:
\begin{equation}
    \label{eq_enc}
    E^{st}_i = \mathbf{Enc^s}(F^{st}_i+\mathbf{PE}(i)),
\end{equation}
where $\mathbf{PE}$ is the vanilla \emph{sinusoidal} position encoding.

The VQ-VAE has demonstrated its effectiveness in generating high-quality musical content \cite{von2023figaro}. Here, we leverage VQ-VAE to focus on generating score-related information at the note level. As shown in Fig.~\ref{xmvae_overall}, the Vector Quantization module maps the encoder output $E^{st}_i$ at each time step to the nearest embedding vector in the codebook $e=\{e_1,\dots, e_K\} \in \mathbb{R}^{d_z \times K}$, where $d_z$ denotes the latent dimension, $K$ is the codebook size. The resulting matched embedding vectors at all time step are represented as the quantized latent sequence $z^s=\{z^s_i\}^T_{i=1} \in \mathbb{R}^{d_z \times T}$, which serves as input to both the Score Decoder and the Performance Decoder. 

Previous studies on decoding compound tokens first introduce a Temporal Decoder ($\mathbf{Dec^s_{tem}}$) to obtain hidden states, followed by various architectures (such as independent linear layers \cite{hsiao2021compound} and simple cross-attention modules \cite{ryu2024nested}) functioning as SubToken Decoder ($\mathbf{Dec^s_{sub}}$) to sample sub-tokens from these hidden states. Different from previous approaches, we propose a Transformer-based SubToken Decoder which effectively decodes sub-tokens of each time step along the sub-token axis. Additionally, our Temporal Decoder is also implemented as a Transformer decoder, operating along the temporal axis. This complementary architecture of decoders shown in Fig.~\ref{xmvae_overall}, processing compound tokens along orthogonal axes, is what we term the orthogonal Transformer decoder. 
Specifically, the embeddings of compound tokens $F^{st}$ and latent sequence $z^s$ from encoder are fed to the Temporal Decoder to obtain the temporal hidden states $H^s$. Then, the SubToken Decoder processes the sub-token embeddings and temporal hidden states $H^s$ and predict the new sub-tokens. The detailed procedure can be formulated as follows:
\begin{equation}
    \label{eq_dec}
    \begin{aligned}
        & H^s_i = \mathbf{Dec^s_{tem}}((F^{st}_i+\mathbf{PE}(i)), z^s), \\
        & \hat{x}^s_{i, j} = \mathbf{Dec^s_{sub}}((F^{st}_{i,j}+\mathbf{PE}(j)), H^s_i), \\
    \end{aligned}
\end{equation}
where $F^{st}_i$ denotes the embeddings of compound token at $i^{th}$ time step, $H^s_i$ is the temporal hidden state at $i^{th}$ time step, serving as the memory for SubToken Decoder, $F^{st}_{i,j}$ is the embedding of $j^{th}$ sub-token at $i^{th}$ time step, and $\hat{x}^{st}_{i,j}$ denotes predicted $j^{th}$ sub-token at $i^{th}$ time step. Note that the inputting embeddings for Temporal Decoder and SubToken Decoder should be right shifted along the temporal axis and the sub-token axis respectively.

After decoding, we can obtain predicted tokens $\hat{X}^{st}=\{\{\hat{x}^{st}_{i,j}\}^{4}_{j=1}\}^{T}_{i=1}$ sampled from the output of \emph{Composer}. 

\subsubsection{Pianist Branch}
Since the quantization of performance parameters are not lossless, we leverage the real-valued expressive parameters $\emph{X}^{pv}$ as depicted in Section~\ref{3_ECP}, to complement the discrete $\emph{X}^{pt}$ as auxiliary inputs for performance modeling. Following the procedure in \eqref{eq_embedding}, we obtain the embeddings $\emph{F}^{pv}$ and $\emph{F}^{pt}$. The embeddings $\emph{F}^{pv}$ and $\emph{F}^{pt}$ are then summed to produce the combined embeddings $\emph{F}^{p}$. Then, $\emph{F}^{p}$ is processed by the Multiscale Performance Encoder, yielding the output $\emph{E}^{p}$ following procedure defined in \eqref{eq_enc}. To capture performance-related information of the entire input sequence, $\emph{E}^{p}$ is averaged along the temporal axis. The averaged output is then passed through two linear layers to compute the mean and variance, which are used to sample the latent variable $z_p \in \mathbb{R}^{d_z}$ of the vanilla VAE. The above procedures are illustrated in the bottom of the Fig.~\ref{xmvae_overall}.

Given that the latent representation produced by the VQ module could be served as high-fidelity features \cite{luo2024bandcontrolnet}, we sum the latent sequence $z^s$ from \emph{Composer} branch with the performance embeddings $F^{pt}$ along the temporal axis. The combined input is then fed into the Performance Decoder, enabling our orthogonal Transformer decoder to leverage score features as guidance for generating the corresponding performance. Following the process similar to \eqref{eq_dec}, the prediction performance tokens $\hat{X}^{p}=\{\{\hat{x}^{p}_{i,j}\}^{4}_{j=1}\}^{T}_{i=1}$ can be sampled from the outputs of the Performance Decoder, where $\hat{x}^{p}_{i,j}$ denotes $j^{th}$ sub-token at $i^{th}$ time step.

\subsubsection{Training and Inference}
The training objective $\mathcal{L}$ of the XMVAE model comprises the VQ-VAE loss $\mathcal{L}^s(\theta)$ in the \emph{Composer} branch and the vanilla VAE loss $\mathcal{L}^p(\phi)$ in the \emph{Pianist} branch. These loss functions are defined as follows:
\begin{equation}
    \label{eq_loss_s}
    \begin{aligned}
        & \mathcal{L}^s(\theta) = & \mathbb{E}_{q_{\theta}(z^s|x^{st})}[-\log p_{\theta}(x^{st}|z^s)] \\
        & & + \alpha \|q_{\theta}(z^s|x^{st})-\mathrm{sg}[e]\|, \\
    \end{aligned}
\end{equation}
\begin{equation}
    \label{eq_loss_p}
    \begin{aligned}
        & \mathcal{L}^p(\phi) = & \mathbb{E}_{q_{\phi}(z^p|x^{pt}, x^{pv})}[-\log p_{\phi}(x^{pt}|z^p, z^s)] \\
        & & + \beta \mathrm{KL}(q_{\phi}(z^p|x^{pt}, x^{pv})\|p_{\phi}(z^p)), \\
    \end{aligned}
\end{equation}
\begin{equation}
    \label{eq_loss_total}
    \mathcal{L} = \mathcal{L}^s(\theta) + \mathcal{L}^p(\phi),
\end{equation}
where $\theta$ and $\phi$ denote the parameters of the respective branches, $\mathrm{sg}[\cdot]$ represents the stop-gradient operator, $\alpha$ and $\beta$ are loss weights, and $\mathrm{KL}$ refers to the Kullback-Leibler divergence. Note that the \emph{Composer} branch is trained without the auxiliary codebook loss, the codebook $e$ is updated using exponential moving averages (EMA) instead \cite{oord2018representation}.

In addition, we train a Transformer decoder-based autoregressive model as the prior model for the VQ-VAE, which learns the categorical distribution $p_{\psi}(z^s) = \prod_i p_{\psi}(z^s_i | z^s_{<i})$ over the quantized embeddings $z^s$ of \emph{Composer} branch.

During inference, the prior model is used to generate a new latent sequence ${z^s}^{\prime}$ autoregressively, while ${z^p}^{\prime}$ is randomly sampled from a normal distribution. The latent sequence ${z^s}^{\prime}$ is fed into the Score Decoder to generate score sub-tokens, while ${z^s}^{\prime}$ and ${z^p}^{\prime}$ is fed into the Performance Decoder to generate token sequences representing expressive parameters. Finally, the token sequences from the two branches are combined and converted into music.

\section{Experiments}
\label{4_experiments}
\subsection{Experiment Settings}
\subsubsection{Dataset}
\label{4_1_dataset}
Our proposed ECP music representation requires dataset containing note-level aligned score and performance data. Here, we conduct our experiments on a score-performance-aligned subset of the ATEPP dataset \cite{zhang2022atepp}, which, to the best of our knowledge, is the largest aligned Western classical piano performance dataset. 

Since ATEPP is transcribed and aligned algorithmically, the original data include much noise and errors. To ensure quality, we only retain performances with top three alignment rate for each musical score.
After that, we curate a refined subset containing 282 musical scores and 787 corresponding performances, achieving an average alignment rate of 83.4\%. Then, we segment all pieces with a sliding window of 256 notes and a stride of 16 notes. The resulting dataset includes 84,483 segments and we call it as \emph{aligned dataset}. We randomly set aside 10\% of all pieces and their corresponding segments for testing, with the remainder used for model training. 

Given the limited scale of the \emph{aligned dataset}, we also collect an additional 1,189 classical piano score files from Musescore\footnote{\url{https://musescore.com/}}. These additional scores are utilized for pretraining VQ-VAE model in the \emph{Composer} branch, with the goal of enhancing the overall performance of the model. We name these additional scores as \emph{pretrain score dataset}.

\subsubsection{Benchmark Models}
We compare XMVAE model with the following four previous models: 1) \textbf{MT} \cite{huang2018music}: It utilizes a MIDI-like representation and a vanilla Transformer decoder with a relative attention mechanism. 2) \textbf{Transformer-GAN} \cite{muhamed2021symbolic}: Built upon MT, it employs a MIDI-like representation and incorporates GAN framework. 3) \textbf{ECPT}: It adopts our ECP representation and uses the architecture same with MT. 4) \textbf{NMT} \cite{ryu2024nested}: The state-of-the-art architecture for decoding compound tokens. We adopts our ECP here.

\subsubsection{Implementation Details}
All models (including the prior model) are configured with a hidden size and embedding size of 256. The encoder/discriminator layers, decoder layers, attention heads, feed-forward network hidden size, and latent size are set to 6, 6, 8, 1024, and 512, respectively, for all benchmarks (where applicable). The NMT and our proposed XMVAE use the same parameter settings, except that their main/temporal decoder and sub-token decoder have 4 and 2 layers, respectively. The codebook size $K$ is set to 512.
All models are optimized using the Adam optimizer. The learning rate is initially warmed up over the first 10 epochs to a maximum value of $2 \times 10^{-4}$, then linearly decayed to $4 \times 10^{-5}$ by the $100^{th}$ epoch, and ultimately stopped at the $200^{th}$ epoch.
The batch size is set to 16. 
All models are trained and evaluated on a single NVIDIA GeForce RTX 3090 GPU with 24 GB of memory. 
During inference, the length of generated compound tokens is uniformly set to 512, and the top-$k$ sampling strategy is applied.

\subsection{Objective Evaluation}
To quantitatively evaluate all models, we use seven objective metrics used in previous studies \cite{muhamed2021symbolic, huang2020pop}, including three aspects:
1) Pitch-based metrics:
\textbf{UPC} (Unique Pitch Classes) denotes the number of different pitch classes in a piece; 
\textbf{PR} (Pitch Range) measures the difference between the highest pitch and lowest pitch; 
\textbf{APS} (Average Pitch by Semitone) represents the average pitch interval in semitones between two consecutive notes.
2) Rhythm-based metrics for assessment of rhythmic structure \cite{huang2020pop}:
\textbf{DSTD} (Downbeat STD) denotes standard deviation of the time difference in seconds between two adjacent estimated downbeats.
\textbf{DS} (Downbeat Salience) represents the mean of all estimated downbeats' salience.
3) Performance-based metrics for evaluation of expressive timing and dynamics:
\textbf{IOI} (Inter-Onset-Interval) represents the time in seconds between the onsets of two consecutive notes.
\textbf{AVI} (Average Velocity Interval) denotes average velocity interval between two consecutive notes in a musical performance.

These metrics are computed for both Real Music (from the testing dataset) and the music generated by the models. For each model, we generate 1,000 music samples and calculate the average values of these metrics as their final scores. Models with metric scores closer to those of the training dataset are considered to perform better. 

\begin{table}[!t]
\tiny
\caption{Results of model comparison. 
The best values (closest to Real Music) are highlighted in \textbf{bold}, while the second-best values are \underline{underlined}. 
}
\centering
\begin{tabular}{lccccccc}
\toprule
    & \textbf{UPC}  & \textbf{PR} & \textbf{APS} & \textbf{DSTD} & \textbf{DS} & \textbf{IOI} & \textbf{AVI} \\
\midrule
\textbf{Real Music}  & 7.6294  & 53.0510 & 11.9777 & 0.2675 & 0.1650 & 0.1075 & 8.7591\\
\midrule
\textbf{MT}                             & 7.0417  & 45.4150 & 10.3968 & 0.3922 & 0.1087 & 0.1478 & 6.8414 \\
\textbf{Transformer-GAN}                & 7.1858  & 47.3981 & 10.5970 & 0.3568 & 0.1218 & 0.1374 & 7.1886 \\
\textbf{ECPT}                           & 8.0794  & 47.7370 & 10.6552 & 0.2289 & \underline{0.1396} & 0.1430 & 7.6362 \\
\textbf{NMT}                            & \underline{7.9823}  & \underline{49.4570} & \textbf{11.4306} & \underline{0.2378} & 0.1332 & \underline{0.1264} & \underline{8.1963} \\
\textbf{XMVAE}                          & \textbf{7.3679}  & \textbf{56.1370} & \underline{10.9181} & \textbf{0.2486} & \textbf{0.1473} & \textbf{0.1240} & \textbf{8.2825} \\
\bottomrule

\end{tabular}
\label{objective_eval_1}
\end{table}

\begin{table}[!t]
\tiny
\caption{Results of ablation studies. All markings in this table have the same meanings as the Table~\ref{objective_eval_1}.}
\centering
\begin{tabular}{lccccccc}
\toprule
    & \textbf{UPC} & \textbf{PR} & \textbf{APS} & \textbf{DSTD} & \textbf{DS} & \textbf{IOI} & \textbf{AVI} \\
\midrule
\textbf{Real Music}  & 7.6294  & 53.0510 & 11.9777 & 0.2675 & 0.1650 & 0.1075 & 8.7591\\
\midrule
\textbf{XMVAE}                          & \underline{7.3679}  & 56.1370 & \textbf{10.9181} & \underline{0.2486} & \textbf{0.1473} & \underline{0.1240} & \underline{8.2825} \\
\quad \textbf{Pretrained}               & \textbf{7.8343}  & \textbf{54.4640} & 10.3316 & \textbf{0.2507} & \underline{0.1412} & \textbf{0.1112} & \textbf{8.4567} \\ 
\quad \textbf{w/o~Multiscale Encoder}   & 7.9080  & 56.0890 & \underline{10.4801} & 0.2481 & 0.1318 & 0.1310 & 7.7163 \\
\quad \textbf{w/o~SubToken Decoder}     & 8.0627  & 47.8260 & 10.2399 & 0.2383 & 0.1369 & 0.1363 & 7.9572 \\
\quad \textbf{w/o~$X^{pv}$}             & 7.9789  & \underline{55.3210} & 10.2647 & 0.2296 & 0.1325 & 0.1357 & 7.1504 \\
\bottomrule

\end{tabular}
\label{objective_eval_2}
\end{table}
\subsubsection{Comparison with Benchmarks}
The results of the model comparison are presented in Table~\ref{objective_eval_1}. Overall, XMVAE achieves the best performance on six out of seven metrics and ranks second on the remaining one, indicating its superior capability in generating high-quality classical piano performances compared to other models. Looking at the comparison between MT and ECPT, which share the same underlying model architecture but utilize different music representations. The superior performance of ECPT highlights the effectiveness of the proposed ECP representation.

\subsubsection{Ablation Studies}
To evaluate the impact of various components in XMVAE, we conducted a series of ablation studies with the following four configurations: 1) \textbf{Pretrained}: The \emph{Composer} branch of XMVAE is pretrained on the \emph{pretrain score dataset}; 2) \textbf{w/o~Multiscale Encoder}: Only the standard self-attention is used in the encoder block of both branches, without incorporating the beat-level self-attention module; 3) \textbf{w/o~SubToken Decoder}: Only the Temporal Decoder is utilized, and sub-tokens decoded independently using linear layers in both branches; 4) \textbf{w/o~$X^{pv}$}: The encoder block of \emph{Pianist} branch is limited to processing only  performance tokens $X^{pt}$, excluding the real-valued expressive parameters $X^{pv}$.

Table~\ref{objective_eval_2} provides the following three findings: 1) XMVAE with a pretrained \emph{Composer} branch achieves the best performance on all objective metrics except for APS and DS, highlighting the benefits of leveraging large-scale musical score datasets to overcome the limitations of a smaller aligned dataset; 2) Excluding the beat-level self-attention from the encoder block or replacing the Transformer-based SubToken Decoder with simpler linear layers leads to poorer performance on most metrics, demonstrating that the additional modules designed for the encoder and decoder significantly enhance the quality of the generated music; 3) Removing the real-valued expressive parameters $X^{pv}$ degrades performance on several metrics, emphasizing the value of incorporating lossless input features to improve the model's predictions.

\begin{table}[!t]
\tiny
\caption{Subjective listening rest results. The best values are highlighted in \textbf{bold}.}
\centering
\begin{tabular}{lccc}
\toprule
    & \textbf{Coherency} & \textbf{Richness} & \textbf{Overall}\\
\midrule
\textbf{Real Music}         &  3.62$\pm$0.26       &  3.80$\pm$0.28       &  3.72$\pm$0.36       \\
\midrule
\textbf{MT}                 &  2.89$\pm$0.43       &  2.92$\pm$0.42       &  2.94$\pm$0.42       \\
\textbf{Transformer-GAN}    &  3.05$\pm$0.33       &  3.22$\pm$0.23       &  3.12$\pm$0.39       \\
\textbf{ECPT}               &  3.06$\pm$0.23       &  3.12$\pm$0.25       &  3.05$\pm$0.34       \\
\textbf{NMT}                &  2.98$\pm$0.66       &  3.31$\pm$0.45       &  3.14$\pm$0.58       \\
\textbf{XMVAE(pretrained)}  &  \textbf{3.38$\pm$0.43}  &  \textbf{3.57$\pm$0.31}   &  \textbf{3.55$\pm$0.43}   \\
\bottomrule
\end{tabular}
\label{subjective_eval}
\end{table}

\subsection{Subjective Evaluation}
In addition to the objective evaluation, we conducted an online listening test to assess the quality of the generated music. Each participant was required to listen to 30 samples which are grouped into five sets. Each set consisted of one Real Music and five generated samples, one from our pretrained XMVAE model and the others from the four benchmark models. It is worth noting that the generated samples were conditioned on the prime of the first beat of their corresponding score sheet in Real Music. All listening samples are presented in random order, and participants rated them based on the following criteria using a 5-point Likert scale (1=very poor, 5=very good): 1) \textbf{Coherency}: Does the music sound fluent and coherent without unnatural transitions? 2) \textbf{Richness}: Does the music sound interesting and diverse, avoiding monotony? 3) \textbf{Overall}: How would you rate the overall quality of the music?

We recruited a total of 26 subjects (9 female and 17 male) to participate this listening test, and 53.8\% of them have classical piano playing experience. Table~\ref{subjective_eval} (Mean values and 95\% confidence intervals are reported.) shows that our XMVAE outperforms other benchmark models across all criteria. 
However, compared to the real music composed and performed by human, the music generated by all models still exhibits some margins, indicating room for improvement.

\begin{figure}[!t]
\centering
\includegraphics[width=3.2in]{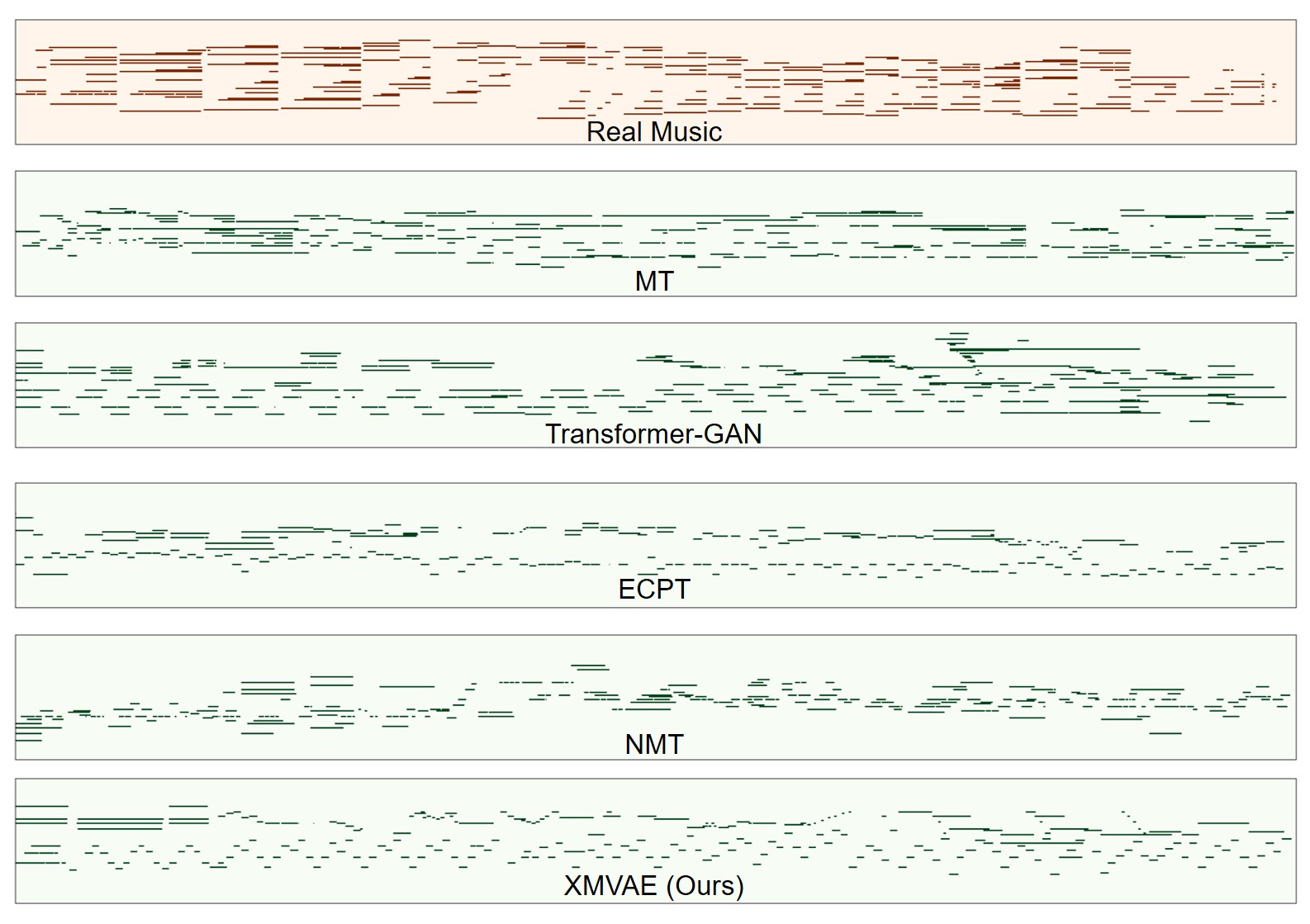}
\caption{Visualized pianoroll of real music and musical samples generated by different models. 
}
\label{case_studies}
\end{figure}

Fig.~\ref{case_studies} showcases one of the five sets used in the listening tests.
The Real Music is characterized by rich chords, a well-defined melodic contour, and a clear metrical structure. The music sample generated by XMVAE demonstrates a clear melodic line supported by concise accompaniment but exhibits engaging development throughout the piece. Conversely, the samples produced by MT, Transformer-GAN, and NMT exhibit poorly aligned melodies and harmonies, resulting in a lack of cohesion and increased dissonance. Meanwhile, the samples from ECPT develop the music with sequences of notes featuring small pitch intervals, which also contribute to a repetitive and less engaging sound.



\section{Conclusion}
\label{5_conclusion}
In this paper, we explore the task of generating classical piano performances from scratch. First, we design the Expressive Compound Word (ECP) representation to faithfully capture both the metrical structure and expressive parameters of classical performances. We then propose the Expressive Music Variational AutoEncoder (XMVAE), which divides the generation process into two branches: a VQ-VAE-based \emph{Composer} branch and a vanilla VAE-based \emph{Pianist} branch. 
Objective and subjective evaluations demonstrate that XMVAE generates classical performances with superior musical quality compared to state-of-the-art models. 
We observe the potential of the XMVAE model to address the task of performance rendering. In this scenario, we can directly feed the latent sequence ${z^s}^{\prime}$ of a given score in to Performance Decoder.
The subsequent steps remain the same as in the generation-from-scratch process. This creative application remains a promising avenue for future exploration. 

\bibliographystyle{IEEEtran}
\bibliography{xmvae.bib}


\addtolength{\textheight}{-12cm}   

\end{document}